\documentclass[aps,prl,twocolumn,groupedaddress,showpacs]{revtex4}
\usepackage[dvipdf]{graphicx}
\usepackage[latin1]{inputenc}
\begin{document}

% Use the \preprint command to place your local institutional report
% number in the upper righthand corner of the title page in preprint mode.
% Multiple \preprint commands are allowed.
% Use the 'preprintnumbers' class option to override journal defaults
% to display numbers if necessary
%\preprint{}

%Title of paper
\title{Glissile dislocations with transient cores in silicon}

% repeat the \author .. \affiliation  etc. as needed
% \email, \thanks, \homepage, \altaffiliation all apply to the current
% author. Explanatory text should go in the []'s, actual e-mail
% address or url should go in the {}'s for \email and \homepage.
% Please use the appropriate macro foreach each type of information

% \affiliation command applies to all authors since the last
% \affiliation command. The \affiliation command should follow the
% other information
% \affiliation can be followed by \email, \homepage, \thanks as well.
\author{Laurent Pizzagalli}
%\email[]{Your e-mail address}
\email{Laurent.Pizzagalli@univ-poitiers.fr}
%\homepage[]{Your web page}
%\thanks{}
%\altaffiliation{}
\affiliation{PHYMAT, Université de Poitiers, CNRS UMR 6630, Bd Marie et Pierre Curie, SP2MI, BP 30179, 86962 Futuroscope Chasseneuil cedex}

\author{Julien Godet}
\affiliation{PHYMAT, Université de Poitiers, CNRS UMR 6630, Bd Marie et Pierre Curie, SP2MI, BP 30179, 86962 Futuroscope Chasseneuil cedex}

\author{Sandrine Brochard}
\affiliation{PHYMAT, Université de Poitiers, CNRS UMR 6630, Bd Marie et Pierre Curie, SP2MI, BP 30179, 86962 Futuroscope Chasseneuil cedex}

\date{\today}

\begin{abstract}

We report an unexpected characteristic of dislocation cores in silicon. Using first-principles calculations, we show that all the stable core configurations for a non-dissociated 60$^\circ$ dislocation are sessile. The only glissile configuration, previously obtained by nucleation from surfaces, surprinsingly corresponds to an unstable core. As a result, the 60$^\circ$ dislocation motion is solely driven by stress, with no thermal activation. We predict that this original feature could be relevant in situations for which large stresses occur, such as mechanical deformation at room temperature. Our work also suggests that post-mortem observations of stable dislocations could be misleading, and that mobile unstable dislocation cores should be taken into account in theoretical investigations. 

\end{abstract}

% 61.72.Lk : Linear defects (dislocations,...)
% 62.20.F- : Deformation and plasticity
% 31.15.E- : Density Functional Theory
% 81.05.Cy : Elemental semiconductors


\preprint{report number}

%\maketitle must follow title, authors, abstract, \pacs, and \keywords
\maketitle

Dislocations are linear defects present in most of materials, and have been largely studied since they are known to strongly influence many properties, primarily mechanical and electrical~\cite{Nab67OUP,Hol07CUP}. Dislocations often play an important role in the plastic deformation of materials, or in epitaxial growth of thin films, because they allow to efficiently relax stress when they move. Mobility is therefore a key property of dislocations, in connection with  macroscopic characteristics such as brittleness or ductility. In a lattice, at rest, dislocations lie in Peierls valleys separated by energy barriers, whose heights define the lattice resistance to dislocation motion~\cite{Hir82WIL}. In materials with deep Peierls valleys, i.e. with moderate to large lattice resistance, such as bcc metals and semiconductors, the properties of dislocations and in particular their mobility typically depend on their core structure, which can be complex and reconstructed~\cite{Cai04DIS}. In order to move, a dislocation must overcome the energy barriers, what can be achieved thanks to the combined action of stress and temperature. 

Considering first the role of temperature, the motion of dislocations is occurring by thermally activated processes such as creation and migration of kinks or jogs. With temperature alone, all possible directions for displacement are equivalent, and a moving dislocation would behave like a random walker. Conversely, in presence of an applied stress, dislocation motion along a specific direction will be favored. In addition,
stress will reduce the energy barriers, thus making easier the dislocation displacement. An important quantity related to dislocation mobility is the Peierls stress, which is the minimum stress required to move a straight dislocation along a given direction without any thermal activation. Dislocations moving in the usual range of temperature and applied stress are called glissile, whereas non-moving dislocations are called sessile.

Theoretical characterizations of dislocation mobility are usually done according to the following procedure. First, low energy configurations for a dislocation core are determined. Then, for the most stable ones, the Peierls stress and the role of thermal activation are investigated. Such an approach is based on the hypothesis that the core configuration for a dislocation at rest is the same than for a moving dislocation. This is a reasonable and general assumption, also employed for point defects diffusion for instance, that has always been verified as far as we know. However, in this letter, we reveal a contradictory situation, where all stable configurations for a dislocation core are sessile, whereas the only glissile configuration is unstable, i.e. not lying in a Peierls valley. Such an unexpected result is fully contrasting with all previous theoretical investigations of dislocation cores, and suggests that it could be advisory to determine mobile dislocation core rather than stable ones. This result also hints that the transient character of some dislocation cores would prevent a direct observation with usual techniques such as post-mortem transmission electron microscopy.

%However, one may wonder whether this hypothesis remains valid in cases for which the stress would be equally or more important than thermal activation.
%remet en cause l'approche "traditionnelle" en simulation des dislocations. En effet, ceci suppose qu'il n'y ait aucune activation thermique, mais de plus chercher un coeur stable ne permet pas de trouver la configuration mobile donc determinante.

\begin{figure}
\includegraphics[width=8.6cm]{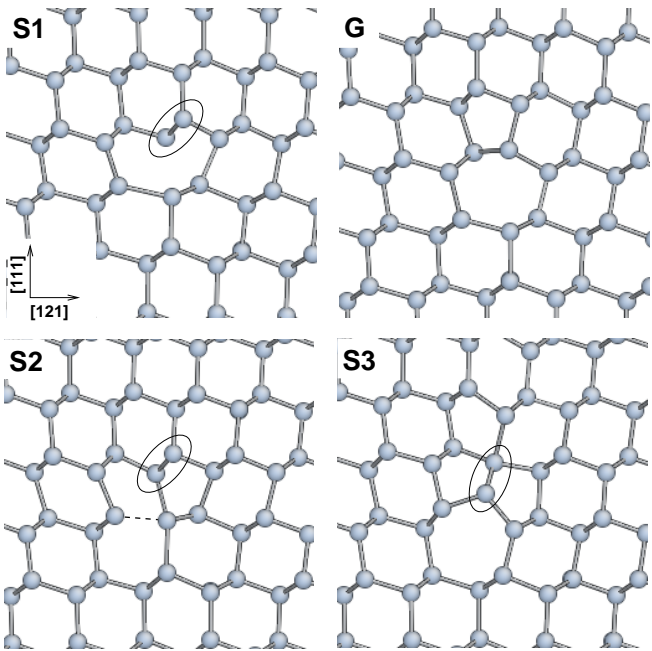}
\caption{Possible atomistic configurations for a 60$^\circ$ dislocation core in silicon, following the notation described in the text. Bonds are drawn according to a criterion of distance between atoms. In the case of the S2 configuration, the dashed line indicates a distance of 2.82~\AA\ between two atoms in the core. An ellipse allows to mark out the two atoms with maximum displacements during the S1$\rightarrow$S3 transformation. }
\label{figstructure}
\end{figure}

In this work, we have used silicon as a model of material with high lattice friction. In its cubic diamond structure, usual dislocations have Burgers vector $\mathbf{b}$ equal to $a_0/2\langle110\rangle$, $a_0$ being the lattice parameter, and orientations screw and 60$^\circ$~\cite{Hir82WIL}. At high temperature ($\agt700$K) and low stress, both are dissociated into 30$^\circ$ or 90$^\circ$ partial dislocations, whereas at low temperature ($\alt700$K) and high stress ($\sim1.5$~GPa) they do not dissociate~\cite{Rab01MSE,Rab07PSS}. In the latter regime, compared to the screw, the 60$^\circ$ dislocation is commonly assumed to be more mobile. As such, it is expected to play an important role for relaxing epitaxial stresses in thin films~\cite{Mar05APL}. Also, it has been shown that the 60$^\circ$ dislocation can form from surface defects~\cite{God04PRB,God06PRB,God09JAP}, a sharp corner~\cite{Izu08JAP}, or in the vicinity of crack fronts~\cite{Alb07perso}, highlighting the fundamental role of this dislocation in the plastic deformation of silicon-based materials. The actual knowledge of the 60$^\circ$ dislocation core geometry is mainly based on an early structure analysis, suggesting two possible core configurations. The first one, proposed by Hornstra~\cite{Hor58JPCS} and called S1 in this work (Fig.~\ref{figstructure}), is located in shuffle \{111\} planes. It is characterized by a 8-atoms ring, containing an atom carrying a dangling bond. Classical potential calculations showed that this dislocation core is stable and that the Peierls stress is about 1.2~GPa~\cite{Li06SM}. This S1 configuration has been identified in previous investigations of the nucleation of dislocations in silicon~\cite{God04PRB,God06PRB,God09JAP,Izu08JAP}. It has also been selected for studying the interaction between dislocation and intrinsic point defects~\cite{Li09SM}. A second possible configuration (called G in Fig.~\ref{figstructure}) is located in glide \{111\} planes~\cite{Hir82WIL}. In that case, the core structure is characterized by two rings of 5 and 7 atoms, and the absence of dangling bonds. Although more stable than the S1 core, this configuration has been shown to be sessile when considered for relaxing epitaxial stresses in thin films~\cite{Mar05APL}.

Using state of the art electronic structure calculations, we have investigated the properties of 60$^\circ$ dislocation, in order to ensure earlier results on the stability and mobility of this dislocation. In a first step, we generated undissociated 60$^\circ$ dislocation core configurations in a silicon cubic diamond crystal using displacements according to anisotropic elasticity theory~\cite{Str58PM}. The dislocation center was set to various positions relatively to the lattice, in order to produce several different dislocation cores. All generated configurations were then relaxed using a classical interatomic potential~\cite{Ter89PRB}. Only two stable dislocation core structures were finally obtained corresponding to the configurations G and S1, shown in the fig.~\ref{figstructure}. Both structures are in agreement with previous works, showing similar topological features as previously depicted.

Next, we investigated the relative stability of both configurations by performing total energy calculations in the framework of density functional theory. Initial computations were made using the DFTB+ code, relying on the tight-binding approximation~\cite{Els98PRB}, and an appropriate basis set~\cite{Sie00THE}. We then employed the first principles SIESTA code~\cite{Sol02JPCM} for obtaining highly accurate results. In the latter case, the local density approximation, norm-conserving pseudoptentials~\cite{Tro91PRB}, and double-$\zeta$ polarized localized orbitals basis set with a cutoff of 6~bohr were used. Optimized lattice constants of 5.46~\AA\ for DFTB+ and 5.404~\AA\ for SIESTA were obtained, in good agreement with the experimental value~\cite{Elasticconst}. Dislocations in bulk silicon were mainly modelled using a quadrupolar arrangement together with periodic boundary conditions, the specific geometry of the cells including only two dislocation cores~\cite{Big92PRL}. Cell dimensions were $12\times12$ along $[121]$ and $[\bar{1}1\bar{1}]$ axis, ensuring that the two dislocation cores were separated by 6 hexagons along $[121]$, and 1 or 2 periods along the dislocation line orientation $[\bar{1}01]$, leading to calculations with either 144 or 288 atoms. Depending on the system size, 1 or 2 k-points along the dislocation line were used. In order to check the possible effect of boundary conditions, test calculations with DFTB+ and clusters including a single dislocation core and at most 232 Si atoms were also performed, surface atoms being fixed and saturated with hydrogen. Both DFTB+ and SIESTA results were fully in agreement, that is why only SIESTA results are described below.

\begin{table}
\caption{Energy differences, for possible core configurations of a 60$^\circ$ dislocation. The most stable reconstructed core configuration G is taken as reference.} \label{nrjtable}
\begin{ruledtabular}
\begin{tabular}{ccccc}
 & S1 & S2 & S3 \\ 
 $\Delta$E (eV/$b$) & $\sim$2.4 & 1.4 & 0.6  \\
\end{tabular}
\end{ruledtabular}
\end{table}

Starting from an initial configuration G, we found very little structural modifications after forces relaxation, performed until all forces were below 10$^{-2}$~eV/\AA. However, we found that the energy of this specific configuration could be further lowered by 0.7~eV/$b$ if the core is reconstructed with a double period along the dislocation line. This reconstructed core has been previously proposed in the case of 60$^\circ$ dislocation in diamond~\cite{Blu02PRB}. In the case of an initial shuffle S1 core, we found the surprising result that this geometry is not stable and transforms to another configuration, approximately 1~eV/$b$ lower, which we call here S2 (fig.~\ref{figstructure}). This unstable behavior, non occuring with classical potentials~\cite{Li06SM}, was obtained in all electronic structure calculations, performed either with DFTB+ and SIESTA. It also did not depend on boundary conditions, since simulations done with cluster systems lead to a similar outcome. We found that it is possible to further decrease the dislocation core energy by slight atomic rearrangements. The new core structure, called S3, is shown in the fig.~\ref{figstructure}, and lowers the energy by 0.8~eV/$b$ compared to the S2 configuration. Compared to the glide core G with the double period reconstruction along the dislocation line, the S3 configuration is 0.6~eV/$b$ higher in energy. The description of the geometry of new configurations S2 and S3, as well as of the transformation mechanisms, is not the focus of this letter, and will be reported elsewhere. Summarizing our results in Table~\ref{nrjtable}, we found two original stable configurations S2 and S3 for shuffle cores, whereas the usually considered geometry S1 is shown to be unstable. The problematics of 60$^\circ$ dislocation in silicon is then more complicated than first thought.

As stated previously, an important property of a dislocation is its mobility, which can be characterized by the Peierls stress. We performed   calculations of the Peierls stress considering all possible stable 60$^\circ$ dislocation configurations. An increasing shear stress, either parallel (defined as positive) or anti-parallel (negative) to the Burgers vector, is applied by straining the computational cell, keeping a constant volume cell, until forces relaxation leads to a displacement of the dislocation. Dislocation cores not displaced for shear strains larger than 15\%, i.e. close to the ideal shear strain~\cite{Rou01PRB,Ume08PRB}, are considered sessile. Note that parallel and anti-parallel shear stress orientations are not equivalent for a 60$^\circ$ dislocation in the cubic diamond structure, yielding two values for the Peierls stress. First, for shear strains as large as 15\% and for both orientations, we found that there is no displacement of the glide core G. This configuration is therefore sessile, in agreement with previous investigations~\cite{Mar05APL}. No attempts were made for determining the shear stress of the double period glide core, but it is very likely that this highly stable and reconstructed core is even more difficult to move. Then, we considered both S2 and S3 shuffle cores. The former is transformed to a S3 configuration for an applied shear strain of -2\% (1.1~GPa), whereas it remains still for all positive strains. The S3 configuration is also not displaced over all the strain range, which is not surprising since it corresponds to a very stable and highly reconstructed core. Additional finite temperature calculations, to be reported elsewhere, indicated that the thermally activated motion of both G and S3 dislocation cores was not favored. 

In previous works, it was commonly assumed that 60$^\circ$ dislocations could exist in two stable configurations, G and S1, with only the latter being glissile. Our simulations rather indicate that the shuffle core S1 is unstable, and that it can be transformed in two different structures, S2 and S3, both being sessile. As a matter of fact, we found no stable and glissile 60$^\circ$ dislocation cores. It is a surprising result, because it is difficult to imagine dislocation loops in silicon with no mobile 60$^\circ$ segments. Furthermore, previous investigations of dislocation nucleation from silicon surfaces, made with first principles calculations, point at formation and propagation of 60$^\circ$ dislocations, with core structures seemingly close to a S1 configuration~\cite{God06PRB}. This last result raises the question of the existence of dislocation cores which could be mobile albeit unstable. To tackle this issue, we tried to determine the stress required for displacing the unstable S1 core in the silicon lattice. We used an initial S1 core obtained as a stable configuration in classical potential calculations, which was subsequently first strained, then relaxed with first principles simulations. For applied strains lower than 2\%, we found that the S1 core evolves to the S2 or S3 configurations. However, for larger strains, typically between 2\% and 5\%~\cite{Strainnote} (1.1 to 2.8~GPa), the S1 core is displaced until it encounters the other dislocation in the cell. During its displacement, the geometry of the S1 core is approximately preserved. Additional calculations have been performed using cluster cells and DFTB+, leading to similar results. This result is in agreement with previous investigations~\cite{God06PRB}. Our conclusion is that only an unstable 60$^\circ$ dislocation core is mobile. This glissile core could therefore be described as transient, since it exists only in motion.

From our calculations, we propose that 60$^\circ$ dislocations that are assumed to participate to the deformation of bulk semiconductors in the high stress/low temperature regime, or to the relaxation of large stresses in epitaxial films, or to the nucleation of dislocations from surfaces, are characterized by this transient core. Typically, there is first nucleation of a 60$^\circ$ dislocation in the unstable mobile configuration followed by the immediate glide of the dislocation, if the applied stress is larger than the Peierls stress. If the stress decreases below this threshold, the dislocation will stop moving and will relax to the sessile core configuration. Such a scenario would occur for any reasons that put the dislocation motionless, even during a very short time. This result, in agreement with previous works showing the nucleation and propagation of the 60$^\circ$ dislocation in silicon~\cite{God06PRB}, has several important and unusual implications. First, the stress required to move a 60$^\circ$ dislocation should always be larger than the Peierls stress. As a consequence, there is no thermally activated motion of the 60$^\circ$ dislocation, even at finite temperature. Second, the role of dislocation nucleation is essential, since it controls the whole process. Finally, since 60$^\circ$ dislocations coming to a halt become sessile, new dislocations have to be formed from other sources for relaxing excess or additional stresses.

In a more general perspective, here we show that mobile dislocation cores, allowing to relax mechanical stresses, can be intrinsically unstable. Consequently, they can hardly be observed experimentally, and especially not in ex-situ experiments. It also raises issues regarding the theoretical investigation of dislocation stability and mobility. In fact, usually one first determines the stable configuration of the dislocation in the Peierls valleys, then after how this configuration would move as a function of stress and temperature. Instead, in our case, dislocation core mobility, and not stability, should be investigated first. Note that it could be a difficult or even unfeasible task. Here, it was fortunate that classical potential calculations spuriously allow to stabilize the mobile core. 

Finally, it is interesting to discuss if this result is a rarity, specific to silicon, or whether it could occur in other systems. Certainly, an important condition is that several core configurations should exist for a given dislocation. This is certainly true in materials where reconstruction of the dislocation core could occur, leading to several more or less complex and stable configurations. All materials showing a covalent bonding character could therefore be concerned: usual semiconductors (Group IV, III-V, or II-VI) or more complex systems such as minerals present in the Earth mantle, or ceramics. Also, stress should be the main cause for dislocation motion compared to thermal activation, suggesting that transient glissile cores could occur in the high stress / low temperature regime. Plastic deformation of several compound semiconductors in this regime has been shown to be similar to what is known for silicon~\cite{Suz98PML,Suz99PMA}. Transient glissile dislocations could also be of prime importance for two important domains of material sciences. The first concerns the onset of plasticity in materials with one or more nanometric dimensions. In such systems, in which there are often no pre-existing defects, and dislocation multiplication is hindered, a large amount of stress can be stored in thin films, nanopillars or nanowires during mechanical testing at room temperature. After nucleation of a primary dislocation from surfaces, dislocation avalanches lead to sudden and large relaxation of stress. 
It is interesting to note that these stresses are typically a significant fraction of the ideal strength of the bulk material, thus large enough for transient glissile core to play a role. As reported previously, such an occurence has been shown for a silicon thin film under large stress~\cite{God06PRB}. We believe that this scenario could also occur in nanowires~\cite{Kiz05PRB,Kan07PM,Zhu09MRSB,Par09MRSB}. The second situation concerns brittle-to-ductile transition, for which dislocation nucleation in the vicinity of the crack front is a critical factor. Here again, large stresses are present, and the occurence of transient glissile dislocation cores should be considered. 

\begin{acknowledgments}
We thank Dr. L.~Kubin for fruitful discussions about the issue of a mobile transient dislocation core. We are also deeply indebted to Dr. T. Albaret for providing us results of Learning-On-The-Fly simulations~\cite{Csa04PRL} revealing the nucleation of 60$^\circ$ dislocations from a crack front in silicon.   
\end{acknowledgments}


\begin{thebibliography}{34}
\bibitem[{Ela()}]{Elasticconst}
\bibinfo{note}{We also computed elastic constants within the same framework. We
  determined the bulk modulus B$=105$~GPa, and elastic constants
  C$_{11}=170$~GPa, C$_{12}=72$~GPa, and C$_{44}=73$~GPa, in fair agreement
  with experimental data~\cite{Hir82WIL}. Our values correspond to a
  $\langle110\rangle$\{111\} shear modulus of 57~GPa.}
\bibitem[{Str()}]{Strainnote}
\bibinfo{note}{We were not able to give more accurate values of the critical
  shear strain required for displacing the 60$^\circ$ dislocation in the S1
  configuration, because of the possible effects due to asymetry of the
  dislocation core, and the use of a core structure previously relaxed with
  classical potentials as initial configuration.}
\end{thebibliography}
\end{document}